\documentclass[12pt]{book}

\usepackage[dvips]{graphicx,color}
\usepackage{makeidx,phisics,cosmology}
\makeauthorindex
\makeindex

\BookTitle{Frontier in Astroparticle Physics and Cosmology}
\CopyRight{\copyright 2004 by Universal Academy Press, Inc.}

\begin{document}

\BookTitle{\itshape Frontier in Astroparticle Physics and Cosmology}
\CopyRight{\copyright 2004 by Universal Academy Press, Inc.}
%\tableofcontents
\pagenumbering{arabic}

\chapter{%   %%%%%%%%% <===== TITLE of the contribution
%%%%%%%%%%% The first letter of each word should be captital letter.
  Pulsar velocities and dark matter hint at a singlet
  neutrino }

\author{%
Alexander Kusenko \\
{\it
Department of Physics and Astronomy, UCLA, Los Angeles, CA
90095-1547 } \\
{\it and RIKEN-BNL Center, Brookhaven National
Laboratory, Upton, NY 11973}
}
%
% Please note:
% One \AuthorContents{} is necessary
% for EACH CONTRIBUTION, for the contents page and
% One \AuthorIndex{} is necessary
% for EACH AUTHOR, for the index.
%
\AuthorContents{A.\ Kusenko} %%%%%%% <=== It is the data for
%%%%%%% CONTENTS. Please enter all author's name that should be
%%%%%%% inithialized. 
\AuthorIndex{Kusenko}{A.} %%%%%%% <=== It is the data for AUTHOR
%%%%%%% INDEX. Please enter a author's name that should be inithialized. 

\section*{Abstract}
Two astrophysical puzzles, the origin of pulsar velocities and that of dark
matter, may have a simultaneous explanation if there exists a sterile
neutrino with a mass in the 1--20 keV range and a small mixing ($\sin
\theta \sim 10^{-4}$) with the electron neutrino.  Although the mixing is
too small for direct detection, future observations of the X-ray
telescopes, as well as the gravity waves detectors, such as LIGO and LISA,
may be able to confirm or rule out the existence of such a neutrino.

\section{Introduction} 

Pulsar velocities range from 100 to 1600~km~s$^{-1}$~\cite{astro,
astro_1}. They present a long-standing puzzle.  According to numerical
simulations of supernova explosions~\cite{explosion}, the asymmetries in
the core collapse could not account for a kick velocity of more than
300-600~km~s$^{-1}$.  Although the average pulsar velocity is in this
range~\cite{astro}, there is a substantial population of pulsars with
velocities in excess of 700~km/s, while as many as 15\% of all pulsars have
speeds over 1000~km~s$^{-1}$~\cite{astro_1}.

It has been suggested that an asymmetric emission of electroweak-singlet
neutrinos from a cooling neutron star could explain the pulsar
velocities~\cite{ks97,fkmp}. Since most of the supernova energy is carried
away by neutrinos, only a few per cent asymmetry is sufficient.  The range
of parameters (mass and mixing angle) consistent with this explanation
overlaps with the allowed region for the singlet neutrinos to make up the
cosmological dark matter~\cite{fkmp,Fuller,dh}.

Let us consider a singlet (or sterile) neutrino $\nu_s$ that has a small
mixing with the electron neutrino $\nu_e$.  The mixing angle $\theta $ we
will consider is $10^{-5} - 10^{-4}$, much smaller than those accessible to
laboratory experiments.  However, neutrino oscillations could result in a
production of such a neutrino in the early universe, as well as in a
supernova.

There are actually two possible mechanisms for the pulsar kick due to the 
sterile neutrinos.  One is based on off-resonance production and requires
the singlet neutrino to have a mass of a few keV~\cite{fkmp}.  The other
possibility is a resonant $\nu_e \rightarrow \nu_s$ conversion, which can
occur for the mass of several to 20 keV. Let us briefly discuss both of
these possibilities.  

\section{Off-resonance emission.} 
In a strong magnetic field inside the neutron star, the electroweak (urca)
processes produce neutrinos with a sizable asymmetry.  However, the
asymmetry in {\em production } does not cause an asymmetry in {\em emission
} of ordinary, active neutrinos because numerous re-scatterings of the
trapped neutrinos wash out the asymmetry~\cite{eq}.  However, if the urca
processes produced some other particles, with a very small scattering cross
section, these particles could escape from the neutron star with an
asymmetry equal to their production asymmetry, hence giving the neutron
star a recoil.  This is what happens if a sterile neutrino exists and if it
has a small mixing with the electron neutrino~\cite{fkmp}.

For a sufficiently small mixing angle between $\nu_e$ 
and $\nu_s$, only one
of the two mass eigenstates,  $\nu_1$, is trapped.  The
orthogonal state, 
$| \nu_2 \rangle = \cos \theta_m | \nu_s \rangle + \sin \theta_m | \nu_e
\rangle , $
escapes from the star freely.  This state is produced in the same basic
urca reactions ($p+e^- \rightleftharpoons \nu_e+n $ 
and
$n+e^+ \rightleftharpoons \bar\nu_e+p$) with the effective Lagrangian
coupling equal the weak coupling times $\sin \theta_m$.  The region of
parameters consistent with the pulsar kick and also with the existing
constraints is shown as region ``2'' in Fig.~1~\cite{fkmp}.

\begin{figure}[t]
  \begin{center}
    \includegraphics[height=9.8cm]{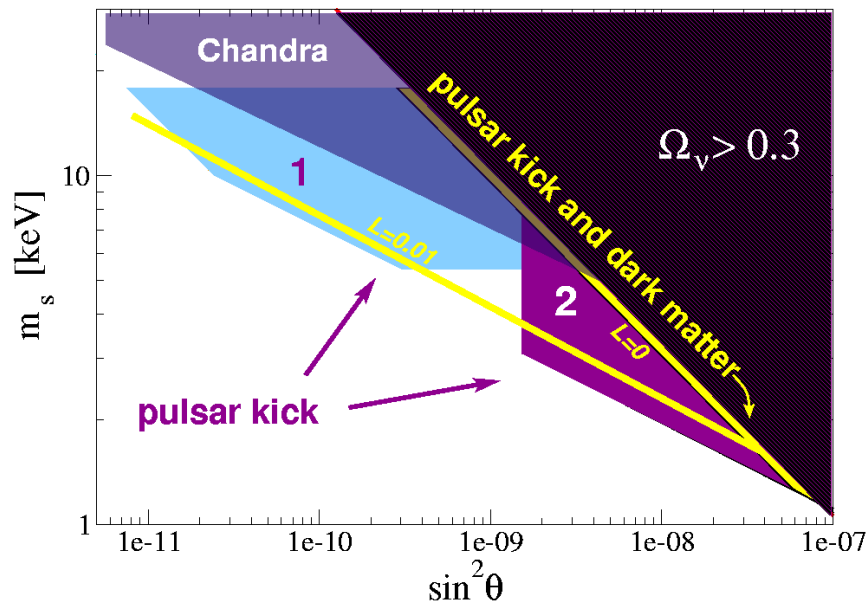}
  \end{center}
  \caption{The region of parameters consistent with the pulsar kicks and
  dark matter. Also shown is the region accessible to Chandra~\cite{aft}.
  In region 1 the pulsar kick is due to the resonant conversions
  $\nu_e\rightarrow\nu_s$~\cite{ks97}.  In region 2, the production of
  singlet neutrinos off resonance can explain the observed pulsar
  velocities~\cite{fkmp}.  The lepton asymmetry $L$ of the universe may affect
  the allowed parameter range for dark matter.  Two narrow bands correspond
  to dark matter for $L=0$ and $L=0.01$.}
\end{figure}

\section{On-resonance emission} 

Resonant production of sterile neutrinos can cause the pulsar kick as well,
for a somewhat different range of masses and mixings.  The position of the
resonant conversion $\nu_e \rightarrow \nu_s $ depends on the direction of
the magnetic field relative to the direction of the neutrino momentum.
Therefore, the sterile neutrinos escape from different densities, {\em
i.e.}, from different depths, depending on their direction.  Due to the
temperature gradient in the star, the average neutrino energy varies with
depth, and so the momentum distribution of emitted sterile neutrinos is
anisotropic.  This mechanism was proposed in Ref.~\cite{ks97} and was a
straightforward generalization of the mechanism proposed
earlier~\cite{ks96} for active neutrinos.\footnote{ I note in passing that
the pulsar kick mechanism based on neutrino oscillations~\cite{ks96,ks97}
was criticized erroneously by Janka and Raffelt~\cite{jr}, who got an
incorrect result by neglecting the differences in the neutrino opacities
for different flavors and by assuming that the flux of neutrinos remained
constant through the core and the neutrinospheres.  The latter assumption
is wrong because both the neutrino absorptions~\cite{ks98} and the
$(1/r^2)$ effect of the spherical geometry~\cite{barkovich} result in a
non-constant flux of the outgoing neutrinos.  These two errors of Janka and
Raffelt were pointed out by us~\cite{ks98} and by Barkovich {\em et
al.}~\cite{barkovich}, respectively.}  (The mechanism using the active
neutrinos alone does not work because the required neutrino mass is too
large~\cite{ks96}.)

\section{Prospects for detection}
The mixing angles shown in Fig.~1 are too small for direct detection.
However, there are ways to confirm or rule out this range of the singlet
masses and mixing angles.  Sterile neutrinos can decay into a (lighter)
active neutrino and a photon.  The lifetime of the sterile neutrinos
exceeds the age of the universe by many orders of magnitude, but there is,
nevertheless, a signal that X-ray telescopes can detect~\cite{aft}.
Chandra observations are sensitive to part of the region shown in Fig.~1,
and Constellation-X may be able cover the entire range, or, at least, a
larger part of it.

A pulsar being accelerated by an asymmetric emission of neutrinos
generates gravity waves~\cite{cuesta,loveridge}.  One can think of it as  
a rotating source of a neutrino jet superimposed on a
spherically symmetric distribution of neutrinos and spinning around some
axis.  This rotating jet produces gravity waves, which may be detected in
the event of a nearby supernova~\cite{loveridge}.

To summarize, a singlet neutrino could explain both the pulsar kicks and
the dark matter simultaneously.  Introduction of such a particle is
probably the most economical extension of the Standard Model that makes it
consistent with dark matter.

This work was supported in part by the DOE Grant DE-FG03-91ER40662 and the
NASA grant ATP02-0000-0151.

%%%%%%%%%%%%%%%%%%%%%%%%%%%%%%%%%%
%% thebibliography environment %%
%%%%%%%%%%%%%%%%%%%%%%%%%%%%%%%%%

%%%%%%%%%%

\end{document}